\documentclass[preprint]{aastex63}
\usepackage[T1]{fontenc}

\usepackage{bm}        
\usepackage{amssymb, amsmath}   
\usepackage{mathtools}
\usepackage{cancel}
\usepackage{color}
\usepackage{graphicx}
\usepackage{dcolumn}

\newcommand\aastex{AAS\TeX}

\DeclareMathAlphabet{\mathsfit}{\encodingdefault}{\sfdefault}{m}{sl}
\SetMathAlphabet{\mathsfit}{bold}{\encodingdefault}{\sfdefault}{bx}{sl}
\newcommand{\vect}[1]{\bm{#1}}

\received{****}
\revised{****}
\accepted{****}
\submitjournal{ApJL}

\shorttitle{\aastex\ Model for Particle Acceleration in Relativistic Reconnection}
\shortauthors{Li et al.}

\begin{document}

\title{A Model for Nonthermal Particle Acceleration in Relativistic Magnetic Reconnection}

\correspondingauthor{Xiaocan Li}
\email{Xiaocan.Li@dartmouth.edu}

\author[0000-0001-5278-8029]{Xiaocan Li}
\affil{Dartmouth College, Hanover, NH 03750 USA}

\author[0000-0003-4315-3755]{Fan Guo}
\affiliation{Los Alamos National Laboratory, Los Alamos, NM 87545, USA}

\author[0000-0001-5880-2645]{Yi-Hsin Liu}
\affil{Dartmouth College, Hanover, NH 03750 USA}

\author[0000-0003-3556-6568]{Hui Li}
\affiliation{Los Alamos National Laboratory, Los Alamos, NM 87545, USA}


\begin{abstract}
  The past decade has seen an outstanding development of nonthermal particle acceleration in magnetic reconnection in magnetically-dominated systems, with clear signatures of power-law energy distributions as a common outcome of first-principles kinetic simulations. Here we propose a semi-analytical model for systematically investigating nonthermal particle acceleration in reconnection. We show particle energy distributions are well determined by particle injection, acceleration, and escape processes. Using a series of kinetic simulations, we accurately evaluate the energy- and time-dependent model coefficients. The resulting spectral characteristics, including the spectral index and lower and upper bounds of the power-law distribution, agree well with the simulation results. Finally, we apply the model to predict the power-law indices and break energies in astrophysical reconnection systems.
\end{abstract}


\section{Introduction}
Nonthermal particle acceleration processes during magnetic reconnection are plausible mechanisms responsible for high-energy emissions observed in magnetically dominated systems, such as relativistic jets from gamma-ray bursts (GRBs)~\citep{Zhang2011Internal,McKinney2011} and active galactic nuclei (AGN)~\citep{Giannios2009Fast,Zhang2015Pol,Zhang2018Large}, pulsar wind nebulae (PWNe)~\citep{Uzdensky2014Physical}, and solar flares~\citep{Lin2003RHESSI}. Recently, interest has surged in studying particle acceleration and resulting particle energy spectra, mainly via particle-in-cell simulations~\citep{Guo2014Formation,Sironi2014Relativistic}. These studies have successfully obtained power-law energy spectra $f(\varepsilon) \propto \varepsilon^{-p}$ with spectral index $p$ decreasing with the plasma magnetization $\sigma$ and approaching 1 when $\sigma\gtrsim10$~\citep{Sironi2014Relativistic,Guo2014Formation,Guo2015Efficient,Werner2014Extent}. However, the origin of the nonthermal energy spectrum and how it varies in different situations is still undergoing active debate.

Despite the advances in kinetic simulations~\citep{Hoshino2001Suprathermal,Zenitani2001Generation,Sironi2014Relativistic,Liu2011Particle,Dahlin2014Mechanisms,Guo2014Formation,Guo2015Efficient,Li2015Nonthermally,Li2018Roles}, theoretical explorations on particle acceleration have limited success in predicting the resulting energetic particle spectrum. Although a simple derivation shows $p \sim 1$ when particle escape is ignored~\citep{Guo2014Formation,Guo2015Particle}, numerous simulations have consistently obtained $p>2$, indicating some key physics is still missing \citep{Werner2018Non,Ball2018,Uzdensky2022}. In addition, how the resulting energy spectra vary with the guide field, and how the results from kinetic simulations can be extrapolated to astrophysical scales are not clear.

In this Letter, we present a model for particle acceleration in magnetic reconnection. In contrast to previous studies, we consider particle injection, acceleration, and escape together as key ingredients for understanding the energy spectra and evaluate them in a reconnection system. In a series of fully kinetic simulations, we quantify the acceleration (both first order and second order) and escape processes with their time and energy dependence. We use the model coefficients to predict the nonthermal power-law characteristics (spectral indices, lower and higher bounds of the power-laws) based on a Fokker-Planck approach and find that the prediction matches the simulations very well. While our model is capable of explaining nonthermal acceleration in kinetic simulations in general, it also reveals essential physical factors that determine energetic particle spectrum, making a crucial step toward a comprehensive understanding of those processes in large-scale astrophysical systems.

\section{A Model}
\begin{figure}[htbp]
  \centering
  \includegraphics[width=0.5\linewidth]{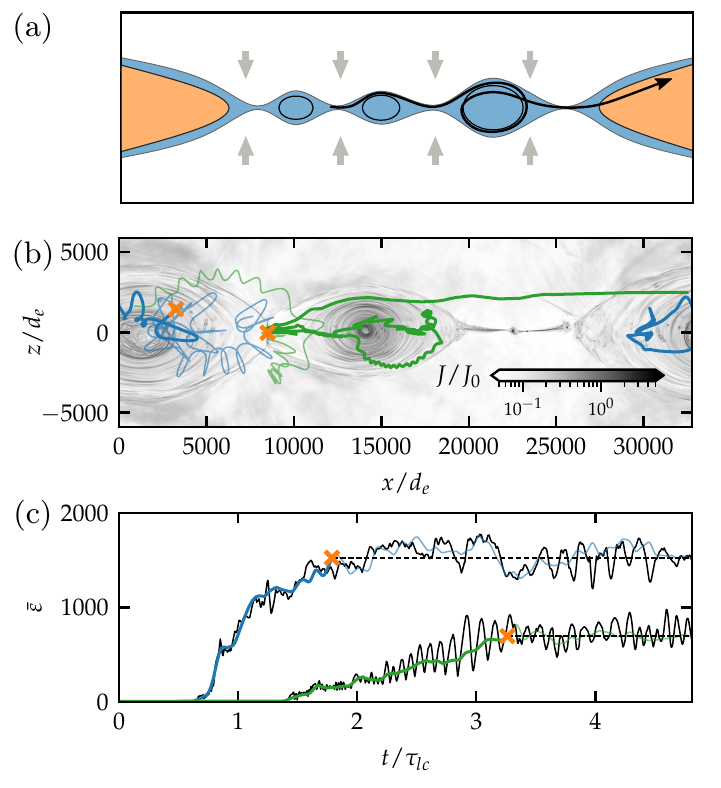}
  \caption{
    \label{fig:cartoon_traj}
    (a) A cartoon illustrating the main processes. The blue and orange regions represent the acceleration and escape regions, respectively. The black curve shows one typical particle trajectory. (b) Two electron trajectories in the kinetic simulations. The orange crosses indicate when particles reach their final energies. The background is at $t\approx2.44\tau_{lc}$, where $\tau_{lc}=L_x/c$. (c) Time evolution of the electron energies. The colored lines are the smoothed data that removed the gyromotion effects.
  }
\end{figure}

Fig.~\ref{fig:cartoon_traj}(a) illustrates our model. The reconnection inflow continuously brings thermal particles into the reconnection layer, where a fraction of them can be accelerated out of the thermal pool through particle injection processes~\citep{Ball2019,Kilian2020Exp,Sironi2022,Guo2023comment,French2023}. A primary acceleration phase then leads to the formation of power-law distribution. While most reconnection studies focus on the first-order acceleration~\citep{Guo2014Formation,Guo2015Particle}, our model includes the second-order Fermi acceleration, which is found to be comparable to the first-order mechanism (see below). This second-order acceleration can be due to inhomogeneous distribution of island acceleration, or turbulent like behavior, similar to relativistic turbulence~\citep{Wong2020First,Comisso2018Particle,Comisso2019Interplay}. In addition, we find that particle escape plays an important role in determining the spectral shape. Since particles do not participate in the acceleration once they are trapped by the largest magnetic islands/flux ropes, they are considered separately as an effective ``escape'' process. Together, we consider these processes in a Fokker-Planck approach~\citep{Blandford1987Particle} describing the evolution of the particle energy distribution $f(\varepsilon)$, according to
\begin{equation}
  \partial_t f + \partial_\varepsilon(\alpha_\text{acc}\varepsilon f) =
  \partial_\varepsilon^2 (D_{\varepsilon\varepsilon} f) - \alpha_\text{esc}f
  + \frac{f_\text{inj}}{\tau_\text{inj}},
  \label{equ:ene_cont2}
\end{equation}

\noindent where $\varepsilon=(\gamma - 1)m_sc^2$ is the kinetic energy, $\alpha_\text{acc}$ is the acceleration rate, $D_{\varepsilon\varepsilon}\coloneqq D_0\varepsilon^2$ is the energy diffusion coefficient, $\alpha_\text{esc}\equiv\tau_\text{esc}^{-1}$ is the escape rate, $f_\text{inj}$ is the injected thermal particle distribution, and $\tau_\text{inj}$ is particle injection time scale. $\alpha_\text{acc}\coloneqq(\partial_t\varepsilon +\partial_\varepsilon D_{\varepsilon\varepsilon})\varepsilon^{-1}$ describes a combination of the first-order Fermi processes and the accompanying first-order term associated with second-order Fermi mechanisms. Instead of separating the injection and nonthermal acceleration, we will treat them as a continuous process and determine $\alpha_\text{acc}$ and $D_0$ due to parallel or perpendicular electric fields ($\vect{E}_\parallel$ and $\vect{E}_\perp$).

We assume that reconnection starts from a single elongated current sheet (with a length of $L_x$), which breaks into a series of magnetic islands (or flux ropes in 3D)~\citep{Loureiro2007Instability,Bhattacharjee2009Fast}. These islands tend to merge to form larger islands, and secondary islands are continuously generated. The entire process lasts a few Alfv\'en-crossing times $\tau_A\coloneqq L_x/V_{Ax}$ until most magnetic flux is reconnected, where $V_{Ax}$ is the reconnection outflow speed. The total number of particles in the reconnection layer due to the reconnection inflow $dN/dt\propto RV_{Ax}$ (where the reconnection rate $R\sim 0.1$~\citep{Liu2017Why,Goodbred2022First}). The largest islands grow in the size of $L_\text{O}=d_i+RV_{Ax}t$, assuming they are initially in the $d_i$ (ion inertial length) scale. For the rest of the discussion, $n_\text{acc}(\varepsilon)$ and $n_\text{esc}(\varepsilon)$ are the numbers of the accelerating and escaped particles in different energy bins, respectively.

For a single electron, the acceleration rate due to the parallel electric field is $-e\vect{E}_\parallel\cdot\vect{v}/\varepsilon$. Since $E_\parallel\approx E_R=RV_{Ax}B_0/c$ near the X-lines and $\approx 0$ rest of the reconnection layer, the acceleration rate due to $\vect{E}_\parallel$ is
\begin{align}
  \alpha_{\text{acc}\parallel} & =\left<-e\vect{E}_\parallel\cdot\vect{v}/\varepsilon\right>, \label{equ:arate_para1} \\
  & \approx ecE_R\left<\mu_E\right>_\text{X}\varepsilon^{-1}F_\text{X}, \label{equ:arate_para2}
\end{align}
where $\left<\cdots\right>$ indicates the ensemble average over $n_\text{acc}(\varepsilon)$ particles, $\left<\cdots\right>_\text{X}$ is the average over $n_\text{X}(\varepsilon)$ particles near the X-lines, and $\mu_E\coloneqq -\vect{v}\cdot\vect{E}_\parallel/(vE_\parallel)$. The filling factor of these electrons $F_\text{X}\sim d_e/L_\text{O}\sim t^{-1}$ for large $t$. $\left<\mu_E\right>_\text{X}\approx 0$ for low-energy thermal electrons with nearly isotropic distributions and increases when they are accelerated by $\vect{E}_\parallel$ or the Fermi mechanism~\citep{Drake2006Electron}. We will determine the exact energy dependence using the simulation results. The energy diffusion rate due to $\vect{E}_\parallel$ is $D_{0\parallel}\coloneqq \left<(\delta\alpha_{\text{acc}\parallel})^2\right>\tau_{\text{dec}\parallel} \approx \left(e^2c^2E_R^2\left<\mu_E^2\right>_\text{X}\varepsilon^{-2}F_\text{X}-\alpha_{\text{acc}\parallel}^2\right)\tau_{\text{dec}\parallel}$, where $\delta\alpha_{\text{acc}\parallel}=-e\vect{E}_\parallel\cdot\vect{v}/\varepsilon-\alpha_{\text{acc}\parallel}$ is the fluctuation of $\alpha_{\text{acc}\parallel}$ and $\tau_{\text{dec}\parallel}\sim d_i/c$ is the decorrelation time for $\vect{E}_\parallel$~\citep[the timescale on which particles see decorrelated $\vect{E}_\parallel$,][]{LeRoux2015Kinetic}. It indicates how fast the broadening of the energy distribution is due to $\vect{E}_\parallel$. Since $\alpha_{\text{acc}\parallel}^2\propto F_\text{X}^2\propto t^{-2}$ for large $t$, the first term dominates. For particles with an anisotropy along the magnetic field, $\left<\mu_E^2\right>_\text{X}$ changes slowly between 0.5 and 1. Therefore,
\begin{equation}
  D_{0\parallel}\approx e^2c^2E_R^2\varepsilon^{-2}F_\text{X}\tau_{\text{dec}\parallel}. \label{equ:d0para}
\end{equation}
The acceleration rate due to $\vect{E}_\perp$ is
\begin{equation}
  \alpha_{\text{acc}\perp} =\left<-e\vect{E}_\perp\cdot\vect{v}/\varepsilon\right>, \label{equ:arate_perp1}
\end{equation}
as particles gaining energy through the Fermi mechanism when colliding with the reconnection outflow~\citep{Guo2014Formation}. The energy gain of each collision is about $\Delta\varepsilon=\left(\Gamma_{Ax}^2\left(1+2V_{Ax}v_x/c^2+V_{Ax}^2/c^2\right)-1\right)\varepsilon$, and the collision time scale is about $L_\text{O}/v_x$, where $\Gamma_{Ax}=(1-V_{Ax}^2/c^2)^{-1/2}$, and $v_x$ is the particle velocity along the reconnection outflow direction. When there is a guide field $B_g=b_gB_0$, $v_x\sim c/(1+b_g)^{1/2}$ on average, and $V_{Ax}=c\sqrt{\sigma_x/(\sigma+1)}$~\citep{Liu2015Scaling}. $\sigma_x=B_0^2/4\pi w$ is plasma magnetization based on the reconnection magnetic field $B_0$~\citep{Liu2015Scaling}, where $w$ is the enthalpy density. $\sigma=B^2/4\pi w$ is total plasma magnetization, where $B=(B_0^2 + B_g^2)^{1/2}$. Thus, $V_{Ax}$ decreases with $b_g$ and increases with $\sigma_x$. We obtain
\begin{align}
  \alpha_{\text{acc}\perp}\sim c\frac{\Gamma_{Ax}^2\left(1+2\beta_{Ax}(1+b_g^2)^{-1/2}+\beta_{Ax}^2\right)-1}{L_\text{O}(1+b_g^2)^{1/2}}, \label{equ:arate_perp2}
\end{align}
where $\beta_{Ax}=V_{Ax}/c$. When the power-law index $p < 2$ in highly magnetized plasmas, most of the kinetic energy is in the high-energy tail~\citep{Sironi2014Relativistic,Guo2014Formation}, and the enthalpy density $w$ keeps increasing. The resulting momentum flux density $w\Gamma_{Ax}^2V_{Ax}^2$ will keep increasing if $V_{Ax}$ stays constant, which is not sustainable. Thus, we expect the reconnection outflow to decrease gradually in highly magnetized plasmas [See Supplemental Material A], resulting in an approximately constant momentum flux density. The energy diffusion rate due to the Fermi mechanism is $D_{0\perp}\sim \Gamma_{Ax}^2\beta_{Ax}^2c\lambda_\text{mfp}^{-1}/3$~\citep{Lemoine2019,Comisso2019Interplay}, where $\lambda_\text{mfp}\sim L_\text{O}(1+b_g^2)^{1/2}$ is scattering mean-free-path. Thus,
\begin{align}
  D_{0\perp}\sim\frac{1}{3}\Gamma_{Ax}^2\beta_{Ax}^2L_\text{O}^{-1}(1+b_g^2)^{-1/2}c.
  \label{equ:d0perp}
\end{align}
Since energetic particles escape the acceleration region when advected by the reconnection outflow, the escape rate is
\begin{align}
  \alpha_\text{esc} & = (dn_\text{esc}/dt) / n_\text{acc}, \label{equ:erate1} \\
  & \sim V_{Ax}/L_\text{O}. \label{equ:erate2}
\end{align}

\section{Numerical Simulations}
We test the proposed model through a series of 2D particle-in-cell kinetic simulations of magnetic reconnection in a proton-electron plasma in the $xz$--plane. The simulations all start from a forcefree current sheet~\citep{Guo2014Formation}. The boundaries along $x$ are periodic for both fields and particles. The boundaries along $z$ are reflective for particles and perfectly conducting for fields. A long-wavelength perturbation with $\delta B_z=0.03B_0$ is included to initiate reconnection~\citep{Birn2001Geospace}. We performed the simulations in proton-electron plasmas ($m_i/m_e=1836$) similar to earlier studies~\citep{Werner2018Non,Kilian2020Exp,Ball2019}. The cold ion magnetization parameters $\sigma_{ic}\coloneqq B_0^2/(4\pi n_im_ic^2)\in[0.4, 1.6, 6.4, 25.6, 102.4]$. The guide fields $b_g\in [0.0, 0.5, 1.0, 2.0]$. The plasmas are initially uniform with density $n_0$ and follow the Maxwell–J\"uttner distributions with a normalized electron temperature $\theta_e=kT_e/m_ec^2=10\sigma_{ic}$ and $T_i=T_e$, resulting in $\sigma_x=0.39, 1.51, 5.2, 13.0, 19.7$ for the runs with different $\sigma_{ic}$. The domain sizes are $L_x\times L_z=L_0\times L_0/2$, where $L_0=4096\sqrt{\theta_e}d_{e0}$ and $d_{e0}\coloneqq c\sqrt{m_e/4\pi n_0e^2}$ is the nonrelativistic electron inertial length. The domain is resolved with a grid $n_x\times n_z=8192\times4096$, and the resulting cell sizes $\Delta x=\Delta z=0.5\sqrt{\theta_e}d_{e0}=0.5d_e$, where $d_e$ is the electron inertial length including the relativistic correction. With $b_g=0.0$, we performed three additional simulations with $L_x\in[L_0/4, L_0/2, 2L_0]$ and the same cell sizes for each $\sigma_{ic}$ to examine the system size dependence. We use 100 particles/cell/species and track about 1 million tracer particles at a high cadence to calculate the model coefficients. These tracer particles are a small, uniformly selected subset of all electrons, and they evolve self-consistently along with the rest of the particles.

\section{Model-Simulation Comparison}
\begin{figure}[ht!]
  \centering
  \includegraphics[width=0.5\linewidth]{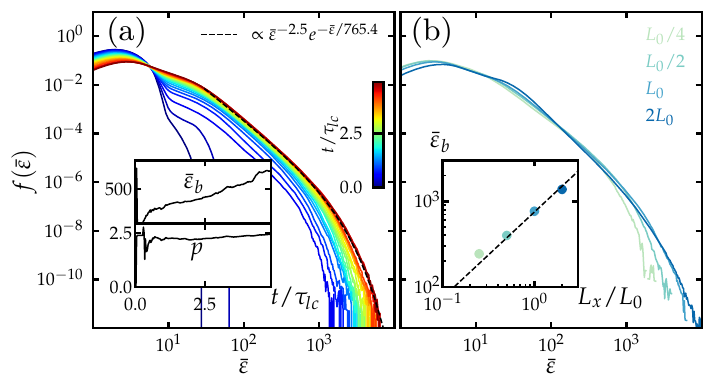}
  \caption{
    \label{fig:spectra}
    Electron energy spectra in the runs with $\sigma_{ic}=6.4$. (a) Time evolution of the spectra when $b_g=0.0$. The insets show the time evolution of the cutoff energy and the power-law index. (b) The spectra in runs with $b_g=0.0$ but different $L_x$. The inset shows the cutoff energy changing with $L_x$. $L_0=4096d_e$ for these runs.
  }
  \end{figure}

\begin{figure*}[ht!]
  \centering
  \includegraphics[width=\linewidth]{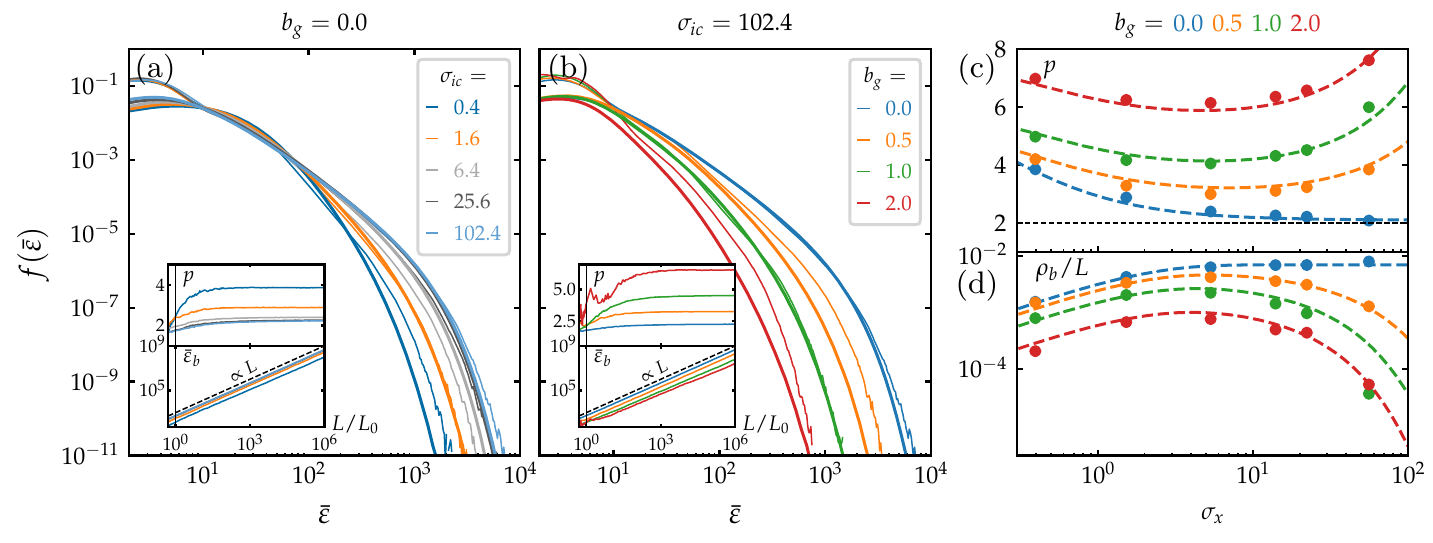}
  \caption{\label{fig:spect_sde}
    (a) Comparing the electron spectra obtained from the model (thick lines) and those in the simulations (thin lines) with $b_g=0.0$ and different $\sigma_{ic}$. (b) Comparing the spectra in the runs with $\sigma_{ic}=102.4$ and different $b_g$. The insets of panels (a) and (b) show the power-law indices and break energies obtained from the models for systems of different sizes. The vertical black line indicate the size of the PIC simulations ($L_0$). (c) Power-law indices in astrophysically relevant systems with different $\sigma_x$ and $b_g$. The dashed are the fittings. (d) The gyroradius of the electrons with the break energies.
  }
\end{figure*}

In all simulations, electron energy spectra develop power-law tails with various power-law indices. Fig.~\ref{fig:spectra}(a) shows a sustainable power-law spectrum with a stable $p=2.5$ and cutoff energy $\bar{\varepsilon}_b$ linearly increasing with time. This result suggests that the cutoff energy increases with the box sizes, confirmed in Fig.~\ref{fig:spectra}(b). As the magnetization increases, the reconnection outflow and the acceleration rate due to the motional electric field will become faster, resulting in a harder spectrum (see Fig.~\ref{fig:spect_sde}). Similarly, the guide field will slow down the outflow and the Fermi acceleration, leading to a softer spectrum (see Fig.~\ref{fig:spect_sde}). We will demonstrate that the model can capture these dependences.

\begin{figure}[ht!]
  \centering
  \includegraphics[width=0.5\linewidth]{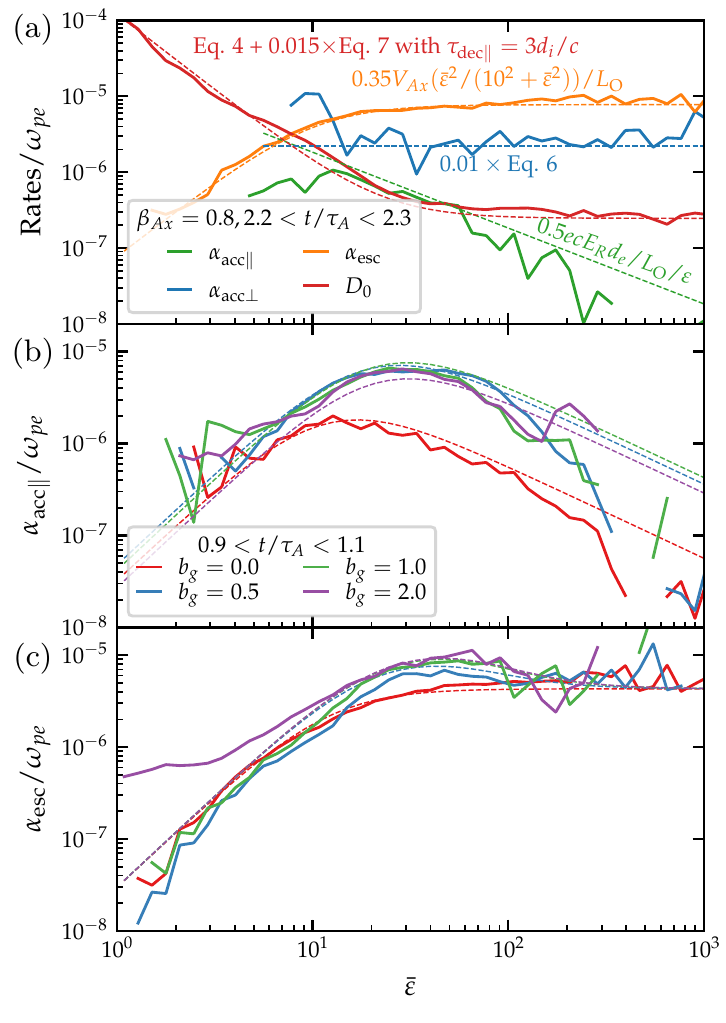}
  \caption{\label{fig:rates_energy}
  The energy dependence of the rates in the runs with $\sigma_{ic}=102.4$. (a) The rates in the run with $b_g=0$. The dashed lines are based on the model scalings. (b) $\alpha_{\text{acc}\parallel}$ for the runs with different guide field. The dashed lines follow the scaling.
  (c) $\alpha_{\text{esc}}$ for these runs.
  }
\end{figure}

\begin{figure}[ht!]
  \centering
  \includegraphics[width=0.5\linewidth]{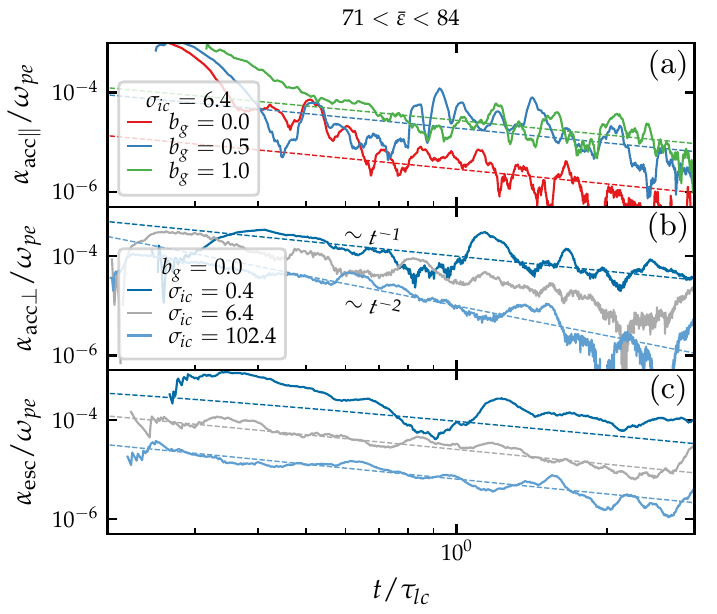}
  \caption{\label{fig:rates_scalings}
    (a) The time dependence of the rates within the selected energy bin for runs with different $b_g$. (b) and (c) The acceleration and escape rates for runs with different $\sigma_{ic}$. The dashed lines follow the predicted scalings in Eqs.~(\ref{equ:arate_para2}),~(\ref{equ:arate_perp2}),~(\ref{equ:erate2}) except in (b), where $\alpha_{\text{acc}\perp}\propto t^{-2}$ when $\sigma_{ic}=102.4$.
  }
\end{figure}

To evaluate the model coefficients in simulations, we must separate particles into the accelerating and the escaped populations. Fig.~\ref{fig:cartoon_traj}(b) \& (c) shows two electron trajectories to illustrate the escape mechanism. The two electrons gain substantial energies before being confined inside the large magnetic island. Their kinetic energies fluctuate without much change after that. Thus, they are treated as escaped particles after their acceleration stops. After separating the two populations, one can calculate acceleration and escape rates using Eqs.~(\ref{equ:arate_para1}),~(\ref{equ:arate_perp1}), and ~(\ref{equ:erate1}). The process is repeated at every step of the tracer particles, allowing us to measure how the rates change over time. We evaluate $D_0$ by tracking the energy spread of the particles in different energy bins~\citep{Comisso2019Interplay,Wong2020First}. Fig.~\ref{fig:rates_energy} shows the energy dependence of the rates in the runs with $\sigma_{ic}=102.4$. $\alpha_{\text{acc}\parallel}$ peaks around $\bar{\varepsilon}=10$ and decreases when $\bar{\varepsilon} > 10$, following the scaling in Eq.~(\ref{equ:arate_para2}). Fig.~\ref{fig:rates_energy}(b) shows that $\alpha_{\text{acc}\parallel}$ peaks at higher energies when $b_g > 0$, due to the increasing $\left<\mu_E\right>_\text{X}$ with $b_g$~\citep{Li2018Roles}. The transition energy $\bar{\varepsilon}_t=\varepsilon_t/(kT_e)\approx 12 + 10\tanh(2b_g)$ according to the simulations, and the dashed lines in Fig.~\ref{fig:rates_energy}(b) follow the energy dependence
\begin{align}
  \alpha_{\text{acc}\parallel}\sim 2.5\times10^{-5}\frac{\bar{\varepsilon}_t^3(\bar{\varepsilon}/\bar{\varepsilon}_t)^2}{\left(1+(\bar{\varepsilon}/\bar{\varepsilon}_t)^2\right)^{3/2}},
  \label{equ:arate_para3}
\end{align}
which is $\sim\bar{\varepsilon}^2$ for low-energy particles and $\sim\bar{\varepsilon}^{-1}$ for high-energy particles (Eq.~\ref{equ:arate_para2}). This energy dependence is determined by fitting the energy-dependent $\alpha_{\text{acc}\parallel}$ to achieve a smooth transition between the two parts. As the rates exhibit significant temporal fluctuations, as illustrated in Fig.~\ref{fig:rates_scalings}, the coefficient should be treated as an approximation rather than an exact value. $\alpha_{\text{acc}\perp}$ is nearly constant, as expected from Eq.~(\ref{equ:arate_perp2}). In Fig.~\ref{fig:rates_energy}(a), we adopt a $\beta_{Ax}=0.8 < \sqrt{\sigma_x/(\sigma+1)}$ due to the decelerating reconnection outflow. Fig.~\ref{fig:rates_energy}(a) shows that while the primary acceleration mechanism at high energies is the Fermi process, the direct and Fermi mechanisms are comparable at low energies, consistent with prior findings by~\citet{Guo2019Determining}. Regarding the escape, $\alpha_\text{esc}$ increases with particle energy until it saturates at a constant $\sim V_{Ax}/L_\text{O}$, consistent with Eq.~(\ref{equ:erate2}).  When $b_g$ is finite, the larger anisotropy indicates that more particles can escape from the acceleration region by streaming along the magnetic field lines. Consequently, $\alpha_\text{esc}$ peaks around $\varepsilon_t$ (Fig.~\ref{fig:rates_energy}(c)), at which the anisotropy is the largest. We use the following function to model the energy dependency of $\alpha_\text{esc}$ from low to high energies.
\begin{align}
  \alpha_\text{esc}\sim 6\times10^{-4}\frac{\bar{\varepsilon}^2\bar{\varepsilon}_t^2}{\bar{\varepsilon}^2+\bar{\varepsilon}_t^2}\left[\left(1-\frac{\bar{\varepsilon}_{t0}^2}{\bar{\varepsilon}_t^2}\right)\frac{10\bar{\varepsilon}_t^2}{10\bar{\varepsilon}_t^2+\bar{\varepsilon}^2} + \frac{\bar{\varepsilon}_{t0}^2}{\bar{\varepsilon}_t^2}\right],\label{equ:erate3}
\end{align}
where $\bar{\varepsilon}_{t0}$ is $\bar{\varepsilon}_{t}(b_g=0)$. $\alpha_\text{esc}\propto\varepsilon^2$ for low-energy particles and approaches $\sim V_{Ax}/L_\text{O}$ for high-energy particles (Eq.~\ref{equ:erate2}). Similar to Eq.~\ref{equ:arate_para3}, the energy dependence is established by fitting energy-dependent $\alpha_\text{esc}$ to achieve a smooth transition between the low-energy and high-energy parts. The complex form accounts for the peak that arises near the transition energy $\varepsilon_t$. Fig.~\ref{fig:rates_energy} (c) shows some discrepancies at low energies, especially for $b_g=2$ (purple curve) case. Such discrepancies are caused by the strong temporal fluctuations in the rates (Fig.~\ref{fig:rates_scalings}), likely due to the initial condition or boundary conditions in the simulations. Fig.~\ref{fig:rates_energy}(a) shows that $D_0$ is dominated by $D_{0\parallel}\propto\varepsilon^{-2}$ (Eq.~\ref{equ:d0para}) at low energies and $D_{0\perp}\sim$ constant (Eq.~\ref{equ:d0perp}) at high energies.

Fig.~\ref{fig:rates_scalings} shows how the rates change with time. Fig.~\ref{fig:rates_scalings}(a) shows that $\alpha_{\text{acc}\parallel}\propto t^{-1}$, following the scaling in Eq.~(\ref{equ:arate_para2}). It is larger when there is a finite guide field, agreeing with Fig.~\ref{fig:rates_energy}(b). Although $\alpha_{\text{acc}\perp}$ strongly fluctuates (Fig.~\ref{fig:rates_scalings}(b)), it decreases with time, as expected from Eq.~(\ref{equ:arate_perp2}). $\alpha_{\text{acc}\perp}$ can be $\propto t^{-2}$ when $\sigma$ is large enough due to the decelerating reconnection outflow [See Supplemental Material A]. Fig.~\ref{fig:rates_scalings}(c) shows that $\alpha_\text{esc}\sim t^{-1}$, consistent with Eq.~(\ref{equ:erate2}). For more model-simulation comparison of the time dependence, see Supplement Material B.

We then solve the Fokker-Planck equation using the energy- and time-dependent rates [See Supplemental Material C for details].
Fig.\ref{fig:spect_sde} (a) and (b) show that the modeled spectra (thick lines) agree well with those in the simulations (thin lines). The model can capture the spectra' $\sigma$- and $b_g$- dependences and the power-law break energies $\varepsilon_b$. Despite some notable agreements between the simulation and model results, as depicted by the overlaying orange curves in Fig.~\ref{fig:spect_sde} (a) and the green and blue curves in Fig.~\ref{fig:spect_sde} (b), there exist discrepancies between the simulation results and model results. Such discrepancies can be attributed to the settings of the simulations, such as the initial and boundary conditions, as well as the constrained system sizes. Consequently, the model results are anticipated to reveal the overall patterns but not necessarily replicate the simulation results exactly.

Since the model is simple enough, we can predict the spectra for much larger systems and study their long-term evolution. The insets of panels (a) and (b) show that $p$ gradually increases until saturation at a constant value in large enough systems. Since the guide field plays a more important role in the high-$b_g$ reconnection, it takes a larger box (up to $10^3L_0$ when $b_g=2$) for the spectra to saturate. The model also shows that the break energy linearly increases with the system size up to $10^6$ times the PIC simulation sizes. Thus, the model can predict the spectra in astrophysical reconnection sites. Fig.~\ref{fig:spect_sde} (c) and (d) show the spectral indices and the gyroradius of the break energies, normalized by the system sizes $L$. When $b_g=0$, $p$ approaches 2.1 as the $\sigma_x$ increases, indicating that the power-law extension can keep growing without causing the kinetic energy to diverge. When $b_g$ is finite, the spectra are the hardest when the $\sigma_x$ is between 1 and 10. $p$ can be approximated as
\begin{align}
  p \approx \frac{1}{\sigma_x+0.2(1+\tanh(b_g))} + 0.04\tanh(b_g)\sigma_x + 1.7b_g + 2.1,\label{equ:pindex}
\end{align}
where the first two terms capture the variation with $\sigma_x$ when the guide field is finite. When $b_g=0$, $p\approx(\sigma_x+0.2)^{-1}+2.1$. For a fixed magnetization $\sigma_x$, the spectrum becomes softer as the guide field becomes stronger. Fig.~\ref{fig:spect_sde} (d) shows the gyroradius of the electrons with the break energies ($\rho_b \coloneqq (\varepsilon_b/m_ec^2 + 1)m_ec/eB_0$) is a fraction of the system size $L$. When $b_g=0$, $\rho_b$ increases with $\sigma_x$ and approaches 0.007 times the system sizes at $2\tau_A$. When $b_g$ is finite, $\rho_b$ peaks when $\sigma_x$ is a few, suggesting the acceleration is most efficient when the $\sigma_x$ is between 1 and 10. $\rho_b$ can be approximated as
\begin{align}
  \frac{\rho_b}{L}=0.007(1-e^{-0.6\sigma_x})e^{-0.06\tanh(b_g)\sigma_x}(1+b_g^2)^{-1}\label{equ:rhob}
\end{align}
at $2\tau_A$. When $b_g=0$, $\rho_b/L=0.007(1-e^{-0.6\sigma_x})$. When $b_g$ is finite, $\rho_b$ peaks when the $\sigma_x$ is between 1 and 10. For a fixed $\sigma_x$, $\rho_b$ monotonically decreases with the guide field. Note that Eqs.~\ref{equ:pindex} and~\ref{equ:rhob} rely only on the hot magnetization parameter $\sigma_x$ and the guide field strength $b_g$. Nonetheless, the equations are applicable to proton-electron plasmas under the conditions where $0.1 < \sigma_x < 100$ and electrons are relativistic. Although the equations can be potentially used for pair plasmas as they do not depend on the ion-to-electron mass ratio, they have not been thoroughly verified against fully kinetic simulations using pair plasmas.

\section{Discussion and Conclusions}
The present study builds upon the prior research by~\citet{Guo2014Formation,Guo2015Particle} and makes improvements in the following aspects: firstly, instead of separating the injection and the nonthermal acceleration, it treats them as a continuous process and incorporates the energy-dependence of acceleration rates; secondly, it incorporates 2nd-order acceleration, including those due to the parallel electric field and the Fermi mechanism; thirdly, it accurately evaluates particle escape rate in the simulations and compares it with the model results; fourthly, it extends the model to proton-electron plasmas; and lastly, besides plasma magnetization, it considers the effect of guide fields on particle acceleration. While these processes are more or less studied in previous studies individually, this study is the first attempt to put these processes together in the same model.

Some important physics is still missing in the current model and could be the key to addressing the particle acceleration in the astrophysical reconnection sites. First, previous research has demonstrated that radiative cooling, such as synchrotron or inverse Compton cooling, significantly impacts the acceleration mechanisms and the dynamics of the current layer, even when cooling is only marginally important~\citep{Zhang2018Large,Werner2019MNRAS,Hakobyan2019ApJ,Sironi2020ApJ,Sridhar2021MNRAS}. Second, 3D reconnection can self-generate plasma turbulence~\citep{Daughton2011Role,Guo2014Formation}, which could affect particle transport and acceleration~\citep{Dahlin2015Electron,Li2019Formation,Zhang2021Efficient}, and change particle acceleration and escape mechanisms~\citep{Zhang2021Fast}. However, other studies have shown that 3D physics is not as crucial in the relativistic regime as in the non-relativistic regime~\citep{Guo2021Magnetic}. Our preliminary results indicate that the spectra are not saturated to the end of the simulation due to the limited system sizes of the 3D simulations. Thus, incorporating 3D effects in the model requires more studies using massive 3D kinetic simulations, which are currently not affordable for revealing the scalings in system sizes. Third, the simulation setup, including boundary conditions and initialization~\citep{Ball2018}, may impact the energy dependence and the time evolution of the rates, as shown in Figs.~\ref{fig:rates_energy} and~\ref{fig:rates_scalings}. These could lead to the deviation of the modeling predictions from the simulation results, given that the semi-analytical model still needs empirical fitting of the kinetic simulation results. These simulations are essential in determining the energy dependencies at low energies and quantifying the constants in each rate for different simulation setups. Addressing all these factors using fully kinetic simulations is critical to making the model more reliable for extending the kinetic simulation results to astrophysical reconnection systems.

In this Letter, we presented a model for determining the main characteristics of power-law particle energy spectra in magnetic reconnection in magnetically dominated plasmas. The power-law spectrum produced by this model is controlled by particle injection, acceleration, energy diffusion, and escape processes. Using a series of first-principles 2D kinetic simulations, we evaluate these transport coefficients and obtained nice agreements with simulation results. By solving the Fokker-Planck equation describing these processes using the modeled transport coefficients, we demonstrate that the modeled spectra agree well with the simulation results for a broad range of magnetization and guide fields. We then use the simulation-verified model to predict the power-law spectral indices and break energies for astrophysically relevant reconnection systems and provide their empirical expressions for different magnetization and guide fields. Our results have strong implications to understanding nonthermal particle acceleration and emissions in high-energy astrophysical systems.

\acknowledgments
We thank the anonymous referee for a constructive review. We gratefully acknowledge our discussions with William Daughton and Patrick Kilian. X.L. acknowledges the support from NASA through Grant 80NSSC21K1313, National Science Foundation Grant No. AST-2107745, and Los Alamos National Laboratory through subcontract No. 622828. The simulations used resources provided by the National Energy Research Scientific Computing Center (NERSC, a U.S. Department of Energy Office of Science User Facility at Lawrence Berkeley National Laboratory) and the Texas Advanced Computing Center (TACC, at the University of Texas at Austin).

\appendix

\section{Decelerating Reconnection Outflow}
\begin{figure}[ht!]
  \centering
  \includegraphics[width=0.5\linewidth]{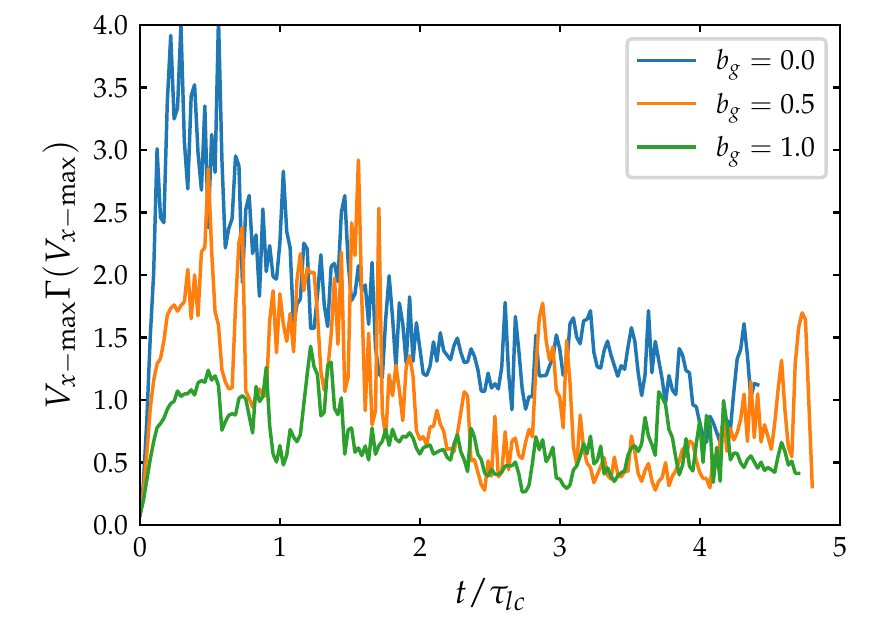}
  \caption{
    \label{fig:vx_gvx}
    Time evolution of $V_{x-\text{max}}\Gamma(V_{x-\text{max}})$ in simulations with $\sigma_{ic}=102.4$, where $V_{x-\text{max}}$ is the maximum outflow speed at the mid-plane and $\Gamma(V_{x-\text{max}})$ is the corresponding Lorentz factor. $\tau_{lc}$ is the light crossing time.
  }
\end{figure}

Fig.~\ref{fig:vx_gvx} shows that the reconnection outflow is decelerating as reconnection evolves. The deceleration is the strongest when the guide field is weak and the particle acceleration is the most efficient.

\section{Time Dependence of the Rates}
\begin{figure}[ht!]
  \centering
  \includegraphics[width=0.5\linewidth]{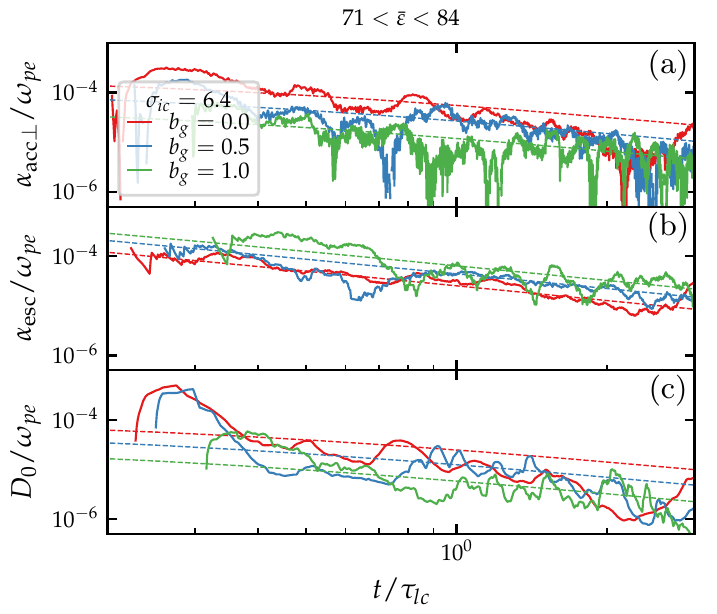}
  \caption{\label{fig:rates_scalings2}
    The time dependence of the rates within the selected energy bin for runs with different $b_g$. The dashed lines follow the predicted scalings $\sim L_\text{O}^{-1}\sim t^{-1}$ in Eqs. (6), (9), and (7).
  }
\end{figure}

Fig.~\ref{fig:rates_scalings2} shows how the rates change with time for runs with different $b_g$. Although $\alpha_{\text{acc}\perp}$ strongly fluctuates (Fig.~\ref{fig:rates_scalings2}(a)), it decreases with time and $b_g$, as expected from Eq. (6) in the main text. When $b_g=0$, due to the decelerating reconnection outflow [See Supplemental Material C], $\alpha_{\text{acc}\perp}$ decreases faster than $t^{-1}$. Fig.~\ref{fig:rates_scalings2}(b) shows that $\alpha_\text{esc}$ is higher when there is a finite guide field. $\alpha_\text{esc}\sim t^{-1}$ (consistent with Eq.~(9) in the main text) even as the reconnection outflow slows down when $b_g=0$ and $\sigma$ is large (Fig.~\ref{fig:rates_scalings2}(f)) because $V_{Ax}$ can be close to $c$ while $\Gamma_{Ax}$ decreases significantly. The energy diffusion rate $D_0$ strongly fluctuates but gradually decreases with $t$. When $b_g=0$, it decreases faster than $t^{-1}$ due to the decelerating outflow.

\section{Solving the Fokker-Planck Equation}
We solve the Fokker-Planck equation~(\ref{equ:ene_cont2})
using the energy- and time-dependent rates. The acceleration part of Eq.~(\ref{equ:ene_cont2}) is equivalent to the stochastic differential equation (SDE) of the It\^{o} type,
\begin{align}
  d\varepsilon=(\alpha_{\text{acc}\parallel}+\alpha_{\text{acc}\perp})\varepsilon dt + \sqrt{2(D_{0\parallel}+D_{0\perp})\varepsilon^2}dW_t, \label{equ:sde}
\end{align}
where $W_t$ is the standard Wiener process, and $dW_t$ is the normalized distributed random number with mean 0 and variance $\Delta t$. The SDE can be solved using pseudo particles. We apply a taper function $\varepsilon_\text{O}/(\varepsilon_\text{O}+\varepsilon)$ to $\alpha_{\text{acc}\perp}$ and $D_{0\perp}$ when solving the equation to model the reduction of the rates with particle energies when their gyroradius is close to the largest island size. Specifically, we use
\begin{align}
  \varepsilon_\text{O}=eB_0\sqrt{1+b_g^2}L_\text{O} / (10m_ec\Gamma_{Ax}),
\end{align}
which includes the correction when the outflow is relativistic. Motivated by the simulation results, we model the reconnection rate $R\approx 0.1V_{Ax}/V_{A0}$, where $V_{A0}=\sqrt{\sigma_x/(\sigma_x+1)}$. As a result, the number of injected particles and the parallel electric field have a strong dependence on the guide field. After updating the energies of the pseudo particles at each time step according to Eq.~(\ref{equ:sde}), we determine whether the particles escape according to the escape rate and inject new pseudo particles. Before the next cycle, we evaluate the enthalpy density of the particles and update the outflow speed $V_{Ax}$ assuming a constant momentum flux density.


\end{document}